\pdfoutput=1
\documentclass[prl,preprint,showpacs,superscriptaddress]{revtex4}
\usepackage{amsfonts,bm}
\usepackage{graphicx}
\usepackage{pstricks}

\newcommand{\be}{\begin{equation}}
\newcommand{\bea}{\begin{eqnarray}}
\newcommand{\ee}{\end{equation}}
\newcommand{\eea}{\end{eqnarray}}
\newcommand{\bpi}{\begin{picture}}
\newcommand{\bce}{\begin{center}}
\newcommand{\epi}{\end{picture}}
\newcommand{\ece}{\end{center}}

\newcommand{\D}{\displaystyle}
\def\chic#1{{\scriptscriptstyle #1}}

\def\gb{\bm{\Gamma}}
\def\kb{\bm{\mathrm K}}
\def\l{\bm{\mathrm L}}
\def\g{\widetilde\gb}
\def\k{\widetilde\kb}

\begin{document}

\title{Gluon and ghost propagators in the Landau gauge:\\ 
Deriving lattice results from Schwinger-Dyson equations}

\author{A.~C. Aguilar}
\affiliation{Departamento de F\'\i sica Te\'orica and IFIC, Centro Mixto, 
Universidad de Valencia-CSIC,
E-46100, Burjassot, Valencia, Spain}

\author{D.~Binosi}
\affiliation{European Centre for Theoretical Studies in Nuclear
Physics and Related Areas (ECT*), Villa Tambosi, Strada delle
Tabarelle 286, I-38050 Villazzano (TN), Italy}

\author{J. Papavassiliou}
\affiliation{Departamento de F\'\i sica Te\'orica and IFIC, Centro Mixto, 
Universidad de Valencia-CSIC,
E-46100, Burjassot, Valencia, Spain}

\begin{abstract}

We show that the application of 
a novel gauge invariant truncation scheme to the Schwinger-Dyson equations
of QCD leads, in the Landau gauge, to an infrared finite
gluon propagator and a divergent ghost propagator, in qualitative 
agreement with recent lattice data.

\end{abstract}

\pacs{
12.38.Lg, 
12.38.Aw,  
12.38.Gc   
}

\maketitle
{\it  Introduction} -- The infrared sector 
of     Quantum     Chromodynamics (QCD)~\cite{Marciano:1977su} 
remains largely unexplored, mainly due to the fact that, unlike 
the electroweak sector of the Standard Model, it 
does not yield to a perturbative treatment.
The  basic building blocks
of  QCD are  the  Green's  (correlation)  functions  of  the
fundamental physical degrees of freedom, gluons and quarks,
and of the unphysical ghosts. 
Even  though it  is well-known
that  these quantities  are not  physical,  since they  depend on  the
gauge-fixing scheme and parameters used  to quantize the theory, 
it is widely believed that reliable information on their non-perturbative
structure is essential for unraveling the infrared dynamics of QCD~\cite{Greensite:2003bk}.

The two basic non-perturbative tools for accomplishing this task  
are ({\it i}) the lattice,  
where space-time is discretized and  the quantities of interest are 
evaluated numerically, and ({\it ii})  the infinite set of coupled non-linear  integral equations
governing  the dynamics of  the QCD Green's functions, known 
as Schwinger-Dyson equations 
(SDE)~\cite{Dyson:1949ha,Schwinger:1951ex,Bjorken:1979dk}.
 Even though  these  equations  are   derived  by  an  expansion  about  the
free-field  vacuum,  they finally  make  no  reference  to it,  or  to
perturbation theory,  and can be  used to address problems  related to
chiral  symmetry  breaking, dynamical  mass  generation, formation  of
bound      states,     and     other      non-perturbative     
effects~\cite{Marciano:1977su}.
While the lattice calculations are limited by 
the lattice size used  and  the  corresponding extrapolation  of  the numerical results to the 
continuous limit, the fundamental conceptual difficulty  in treating  the SDE  
resides  in the need  for a  self-consistent truncation  scheme,  {\it i.e.},   one   that  
does  not  compromise  crucial properties of the quantities studied.

It it  generally accepted by now  that the lattice  yields in the 
 Landau  gauge  (LG)
an infrared finite  gluon  propagator 
and an infrared divergent ghost  propagator.
This  rather characteristic  behavior has been  firmly established  recently using
large-volume lattices, for pure  Yang-Mills (no quarks included), for
both $SU(2)$~\cite{Cucchieri:2007md} and $SU(3)$~\cite{Bogolubsky:2007ud}.  
To be sure, lattice simulations  of gauge-dependent quantities are known to 
suffer from the problem of the Gribov copies, especially in the infrared 
regime, but it is generally believed 
that the effects are  quantitative rather than qualitative. 
The effects of the Gribov ambiguity on the ghost propagator 
become more pronounced in the infrared, 
while their impact on the gluon propagator 
usually stay within the statistical error of the simulation~\cite{Williams:2003du}.      
In what follows we will assume that in the lattice results 
we use the Gribov problem is under control.


In this article we show that  the SDEs obtained within a new 
gauge-invariant truncation scheme  
furnish results (in the LG) which are in qualitative agreement with the lattice data.
As has been first explained in ~\cite{Cornwall:1982zr}, 
obtaining an infrared finite result for the gluon self-energy from SDEs,
without violating the underlying local gauge symmetry, is far from trivial,
and hinges crucially on one's ability to devise
a  self-consistent truncation  scheme  that would  select a  tractable
and, at the same time, {\it physically meaningful} 
subset of these equations. To accomplish this, in the present work 
we   will   employ   
the  new gauge-invariant  truncation  scheme derived in \cite{Binosi:2007pi},  
which is based  on  the  pinch  technique~\cite{Cornwall:1982zr, Cornwall:1989gv} and  its 
correspondence~\cite{Binosi:2002ez} 
with the background field method (BFM)~\cite{Abbott:1980hw}.

\begin{figure*}
\includegraphics[width=14cm]{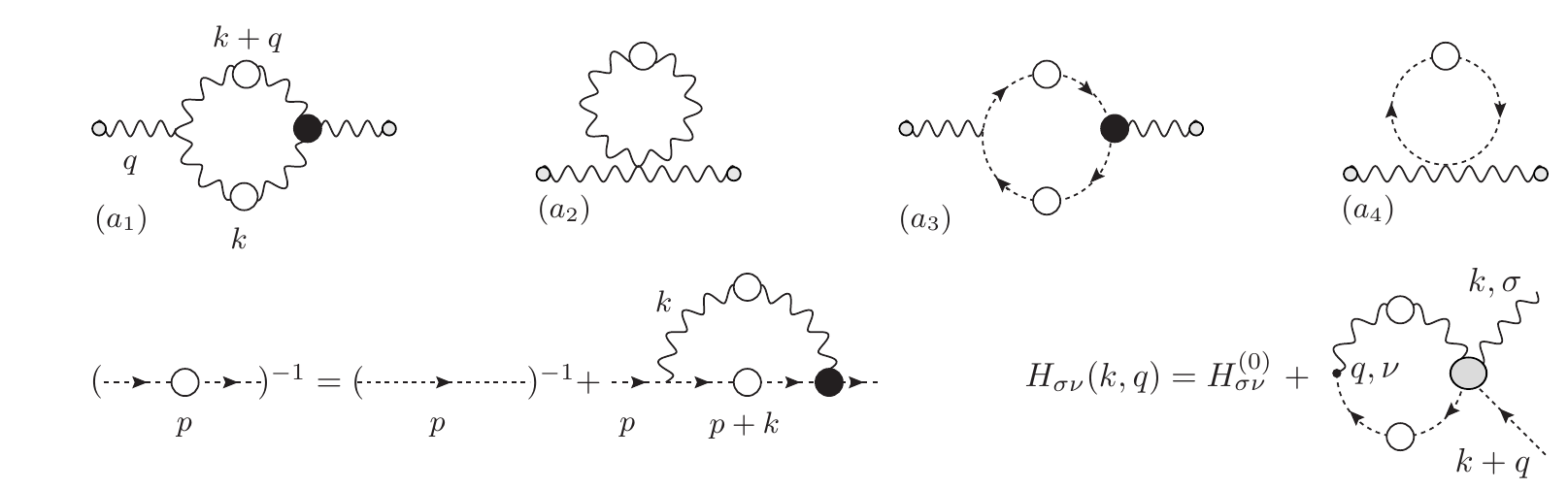}
\caption{The new SDE for the gluon-ghost system.
Wavy lines with white blobs are 
full gluon propagators, dashed lines with 
white blobs are full-ghost propagators, black blobs are  full vertices, and 
the grey blob denotes the scattering kernel. The circles attached 
to the external gluons denote that, 
from the point of view of Feynman rules, they are treated as background fields.}
\label{SDeqs}
\end{figure*}

{\it SDEs in the gauge-invariant truncation scheme} -- 
The gluon propagator $\Delta_{\mu\nu}(q)$ 
in the covariant gauges assumes the form 
\be
\Delta_{\mu\nu}(q)= -i\left[ {\rm P}_{\mu\nu}(q)\Delta(q^2) +\xi\frac{\D q_\mu
q_\nu}{\D q^4}\right],
\label{prop_cov}
\ee
where $\xi$ denotes the gauge-fixing parameter, 
\mbox{${\rm P}_{\mu\nu}(q)= g_{\mu\nu} - q_\mu q_\nu /q^2$}
is the usual transverse projector, and, finally, $\Delta^{-1}(q^2) = q^2 + i \Pi(q^2)$, 
with  $\Pi_{\mu\nu}(q)={\rm P}_{\mu\nu}(q) \,\Pi(q^2)$ the gluon self-energy. 
In addition, the full ghost propagator $D(p^2)$ and its self-energy $L(p^2)$ are related by
$iD^{-1}(p^2)= p^2 - iL(p^2)$. In the case of pure (quarkless) QCD,
the new SD series~\cite{Binosi:2007pi} for the gluon and ghost propagators reads (see also Fig.~\ref{SDeqs})
\bea
&& \Delta^{-1}(q^2){\rm P}_{\mu\nu}(q) = 
\frac{q^2 {\rm P}_{\mu\nu}(q) + i\,\sum_{i=1}^{4}(a_i)_{\mu\nu}}{[1+G(q^2)]^2} \,,
\nonumber\\
&& iD^{-1}(p^2) = p^2 +i \lambda  \int_k
\Gamma^{\mu}\Delta_{\mu\nu}(k)\gb^{\nu}(p,k) D(p+k) \,,
\nonumber\\
&& i\Lambda_{\mu \nu}(q) = \lambda 
\int_k H^{(0)}_{\mu\rho}
D(k+q)\Delta^{\rho\sigma}(k)\, H_{\sigma\nu}(k,q)\,,
\label{newSDb}
\eea
where \mbox{$\lambda=g^2 C_{\rm {A}}$}, with $C_{\rm {A}}$ the Casimir eigenvalue of the adjoint representation
[$C_{\rm {A}}=N$ for $SU(N)$], 
and \mbox{$\int_{k}\equiv\mu^{2\varepsilon}(2\pi)^{-d}\int\!d^d k$}, 
with $d=4-\epsilon$ the dimension of space-time. 
$\Gamma_{\mu}$ is the standard (asymmetric) gluon-ghost vertex at tree-level,
and $\gb^{\nu}$ the fully-dressed one.
$G(q^2)$ is the $g_{\mu\nu}$ component
of the auxiliary two-point function $\Lambda_{\mu \nu}(q)$, and the function $H_{\sigma\nu}$ is 
defined diagrammatically in Fig.~\ref{SDeqs}.
$H_{\sigma\nu}$ is in fact a familiar object~\cite{Marciano:1977su}: 
it appears in the all-order Slavnov-Taylor identity (STI)
satisfied by the standard  three-gluon vertex, and
is related to the full gluon-ghost vertex by 
\mbox{$q^{\sigma} H_{\sigma\nu}(p,r,q) = -i{\gb}_{\nu}(p,r,q)$}; 
at tree-level, $H_{\sigma\nu}^{(0)} = ig_{\sigma\nu}$.

When evaluating  the diagrams $(a_i)$  one should use the  BFM Feynman
rules~\cite{Abbott:1980hw};  notice in particular  that ({\it  i}) the
bare  three- and  four-gluon  vertices depend  explicitly on  $1/\xi$,
({\it ii})  the coupling of  the ghost to  a background gluon  is {\it
symmetric} in  the ghost  momenta, ({\it iii})  there is  a four-field
coupling between two  background gluons and two ghosts.  Thus, for the
gluonic contributions we find
\begin{eqnarray}
&& (a_1)_{\mu\nu} =
\frac{\lambda}{2}\int_k
\widetilde{\Gamma}_{\mu\alpha\beta}
\Delta^{\alpha\rho}(k)
{\g}_{\nu\rho\sigma}
\Delta^{\beta\sigma}(k+q)  \,,
\nonumber\\
&& (a_2)_{\mu\nu} =
-i\lambda g_{\mu\nu}\int_k \!\Delta^{\rho}_{\rho}(k)
-i \lambda \left(\frac{1}{\xi}-1\right) \int_k \!\Delta_{\mu\nu}(k),
\label{groupa}
\end{eqnarray}
with \mbox{$\widetilde{\Gamma}^{\mu\alpha\beta}\!(q,p_1,p_2)\! =\!
\Gamma^{\mu\alpha\beta}\!(q,p_1,p_2)\! + \!(p_{2}^{\beta}g^{\mu\alpha}\!\!-\!p_{1}^{\alpha}g^{\mu\beta})\xi^{-1}\!,$}
$\Gamma_{\mu\alpha\beta}$ the standard QCD three-gluon vertex, and ${\g}_{\mu\alpha\beta}$ is
the fully-dressed version of $\widetilde{\Gamma}_{\mu\alpha\beta}$.
For the ghost contributions, we have instead
\begin{eqnarray}
&& (a_3)_{\mu\nu} = -\lambda \int_k  \widetilde{\Gamma}_{\mu} D(k)  D(k+q) {\g}_{\nu}\,,
\nonumber\\
&& (a_4)_{\mu\nu} =  2 i\lambda g_{\mu\nu} \int_k D(k)\,,
\label{groupb}
\end{eqnarray}
with \mbox{$\widetilde{\Gamma}_{\mu}(q,p_1,p_2) = (p_2-p_1)_{\mu}$}, and ${\g}_{\mu}$ its fully-dressed counterpart.
Due to the Abelian all-order Ward Identities (WIs) that these two full 
vertices  satisfy (for all $\xi$), namely
\mbox{$q^{\mu}{\g}_{\mu\alpha\beta}=i\Delta^{-1}_{\alpha\beta}(k+q)
-i\Delta^{-1}_{\alpha\beta}(k)$} and  \mbox{$
q^{\mu}{\g}_{\mu} = iD^{-1}(k+q) - iD^{-1}(k)$},
one can demonstrate that 
$q^{\mu}[(a_1)+(a_2)]_{\mu\nu} =0$
and \mbox{$q^{\mu}[(a_3)+(a_4)]_{\mu\nu} =0$} \cite{Aguilar:2006gr}.

For the rest of the article we will study the system of coupled SDEs~(\ref{newSDb})
in the LG ($\xi =0$),  in order to make contact with the recent lattice results of~\cite{Bogolubsky:2007ud,Cucchieri:2007md}. 
This is a subtle exercise, because one cannot set directly $\xi =0$ in the integrals on the rhs of (\ref{groupa}), due to the terms proportional to  $1/\xi$. Instead, one
has to use the expressions for general $\xi$, carry out explicitly the set of cancellations produced when the terms proportional to $\xi$ generated by the identity
$k^{\mu} \Delta_{\mu\nu}(k)= -i \xi k_{\nu}/k^2$ are used to cancel $1/\xi$ terms, and set $\xi =0$ only at the very end. 
It is relatively easy to establish that only the bare part 
$\widetilde{\Gamma}_{\nu\alpha\beta}$ of the full vertex 
contains terms that diverge as $\xi\to 0$. 
Writing ${\g}_{\nu\alpha\beta} = \widetilde{\Gamma}_{\nu\alpha\beta} + {\k}_{\nu\alpha\beta}$,
we thus have that ${\k}_{\nu\alpha\beta}$ is regular in that limit, and we will denote by ${\kb}_{\nu\alpha\beta}$ its value at $\xi =0$. 
Introducing $\Delta^\mathrm{t}_{\mu\nu}(q)={\rm P}_{\mu\nu}(q)\Delta(q^2)$, 
we get
\bea
\sum_{i=1}^2(a_i)_{\mu\nu}&=&\lambda\Bigg\{\frac{1}{2}\int_k
\Gamma_\mu^{\alpha\beta}\Delta_{\alpha\rho}^\mathrm{t}(k)\Delta_{\beta\sigma}^{t}(k+q)\l_\nu^{\rho\sigma}
-\frac{9}{4}g_{\mu\nu}\int_k \Delta(k) \nonumber \\
&+&\int_k \!\!\Delta_{\alpha\mu}^\mathrm{t}(k) 
\frac{(k+q)_{\beta}}{(k+q)^2}[\Gamma+ {\l}]_\nu^{\alpha\beta}
+\int_k \frac{k_{\mu}(k+q)_{\nu}}{k^2 (k+q)^2}\Bigg\},
\label{contr}
\eea
where $\l_{\mu\alpha\beta}=\Gamma_{\mu\alpha\beta}+\kb_{\mu\alpha\beta}$ satisfies the WI
\mbox{$q^{\mu} \l_{\mu\alpha\beta} = {\rm P}_{\alpha\beta}(k+q)\Delta^{-1}(k+q)
-{\rm P}_{\alpha\beta}(k)\Delta^{-1}(k)$}. Contracting the 
lhs of (\ref{contr}) by $q^{\mu}$ one can then 
verify that it vanishes, as announced.

Next, following standard techniques,
we express ${\l}_{\mu\alpha\beta}$ and ${\g}_{\mu}$
as a function of the gluon and ghost self-energy, respectively, in such a way as to 
automatically satisfy the corresponding WIs. 
Of course, this method leaves the transverse 
({\it i.e.}, identically conserved) part of the vertex undetermined. 
The Ansatz we will use is 
\bea
{\l}_{\mu\alpha\beta}&=& \Gamma_{\mu\alpha\beta} + i\frac{q_{\mu}}{q^2}
\left[\Pi_{\alpha\beta}(k+q)-\Pi_{\alpha\beta}(k)\right]\nonumber \,,  \\
{\g}_{\mu}&=& \widetilde{\Gamma}_{\mu} -i\frac{q_{\mu}}{q^2}
\left[L(k+q)-L(k)\right]\,,
\label{gluonv}
\eea
whose essential feature is the presence of massless pole terms, $1/q^2$.
Longitudinally coupled bound-state poles    
are known to be instrumental for obtaining 
\mbox{$\Delta^{-1}(0)\neq 0$}~\cite{Jackiw:1973tr}; on the other hand, due to 
current conservation, they do not contribute to the $S$-matrix.
For the conventional ghost-gluon vertex ${\gb}_{\nu}$,
appearing in the second SDE of (\ref{newSDb}) we will use its
tree-level expression, {\it i.e.}, ${\gb}_{\nu}\to \Gamma_{\nu} = -p_{\nu}$.
Note that, unlike ${\g}_{\nu}$,  the conventional ${\gb}_{\nu}$ 
satisfies a STI of rather limited usefulness; the ability to 
employ such a different treatment for ${\g}_{\nu}$ and  ${\gb}_{\nu}$  
without compromising gauge-invariance 
is indicative of the versatility of the new SD formalism used here.
Finally, for $H_{\sigma\nu}$ we use its tree-level value, $H_{\sigma\nu}^{(0)}$.

With these approximations, the last two equations of~(\ref{newSDb}), together
with (\ref{groupb}) and (\ref{contr}), give (in Euclidean space) 
\begin{widetext}
\bea
& & [1+G(q^2)]^2 \Delta^{-1}(q^2)\, =\, 
q^2 - \frac{\lambda}{6} \left[ 
\int_k \!\!\Delta(k)\Delta(k+q)f_1 
+ \int_k  \!\!\Delta(k) f_2 
- \frac{1}{2} \int_k \frac{q^2}{k^2 (k+q)^2}\right] 
\nonumber\\
&&\hspace{3.7cm}\,+\, \lambda \bigg[\frac{4}{3}
\int_k \left[ k^2 - \frac{(k\cdot q)^2}{q^2}\right] D(k) D(k+q)
- 2 \int_k  D(k)\bigg]\,,
\label{sdef} \\
& &f_1 = 20q^2 + 18k^2 -6(k+q)^2 + \frac{(q^2)^2}{(k+q)^2}-  (k\cdot q)^2\bigg[ \frac{20}{k^2}
+ \frac{10}{q^2} + \frac{q^2}{k^2 (k+q)^2}
+\frac{2 (k+q)^2}{q^2 k^2}\bigg] \,,\nonumber\\
& & f_2 = -\frac{27}{2} -8 \frac{ k^2}{(k+q)^2}
+8 \frac{q^2}{(k+q)^2} 
+ 4 \frac{(k\cdot q)^2}{k^2(k+q)^2}
- 4 \frac{(k\cdot q)^2}{q^2(k+q)^2} \,,\nonumber
\label{DSDE}
\eea
\end{widetext} 
\bea
D^{-1}(p^2) &=& p^2 -  \lambda \int_k\,\left[p^2-\frac{(p \cdot k)^2}{k^2}
\right]\Delta(k)\,D(p+k) \,,
\nonumber\\
G(q^2) &=& - \frac{\lambda}{3}\int_k\,\left[
2+\frac{(k\cdot q)^2}{k^2q^2}\right]\Delta(k)D(k+q) \,.
\label{gg}
\eea 
Since \mbox{$[(a_1)+(a_2)]_{\mu\nu}$} and \mbox{$[(a_3)+(a_4)]_{\mu\nu}$} are transverse,  
in arriving at (\ref{sdef}) we have used 
$[(a_1)+(a_2)]_{\mu\nu} = {\rm Tr}[(a_1)+(a_2)] {\rm P}_{\mu\nu}(q)$
and $[(a_3)+(a_4)]_{\mu\nu} = {\rm Tr}[(a_3)+(a_4)] {\rm P}_{\mu\nu}(q)$,
substituted into (\ref{newSDb}), and then equated the scalar co-factors of both sides. 
Thus, the transversality of the answer cannot be possibly compromised by the 
ensuing numerical treatment (e.g. hard ultraviolet cutoffs), 
which may only affect the value of the co-factor.

{\it Numerical results} -- Before solving numerically the above system of integral equations, 
one must introduce renormalization 
constants to 
make them finite. The values of these constants will be fixed by the 
conditions  $\Delta^{-1}(\mu^2)=\mu^2$, $D^{-1}(\mu^2)=\mu^2$, and 
$G(\mu^2)=0$, with the renormalization point  
$\mu^2$ of the order of $M_\chic{Z}^2$.
It is relatively straightforward to verify that 
the perturbative expansion of (\ref{sdef}) and (\ref{gg})
furnishes the correct one-loop results.
Specifically, keeping only leading logs, we have  
\mbox{$1+G(q^2) = 1 +\frac{3C_{\rm {A}}\alpha_s}{16\pi}\ln(q^2/\mu^2)$}, while
\mbox{$D^{-1}(p^2)= p^2 [1+\frac{3C_{\rm {A}}\alpha_s}{16\pi}\ln(p^2/\mu^2)]$} and 
\mbox{$\Delta^{-1}(q^2)= q^2 [1+\frac{13C_{\rm {A}}\alpha_s}{24\pi}\ln(q^2/\mu^2)]$},
where \mbox{$\alpha_s = g^2/4\pi$}.

The crux of the matter, however, is the behavior of 
(\ref{sdef}) as $q^2\to 0$, where the ``freezing'' of the 
gluon propagator is observed.
In this limit, Eq.(\ref{sdef}) yields
\bea
& & \Delta^{-1}(0) = \frac{\lambda \left(T_{g} + T_{c}\right)}{[1+G(0)]^2},
\label{tad}\\
&& T_{g}= \frac{15}{4} \int_k \Delta(k) 
- \frac{3}{2} \int_k k^2 \Delta^2(k),
\label{gltad}
\\
&& T_{c} =-2 \int_k D(k) + \int_k k^2 D^2(k).
\label{ghtad}
\eea

\begin{figure*}[t]
\centering
\begin{tabular}{cc}
\includegraphics[width=8.1cm]{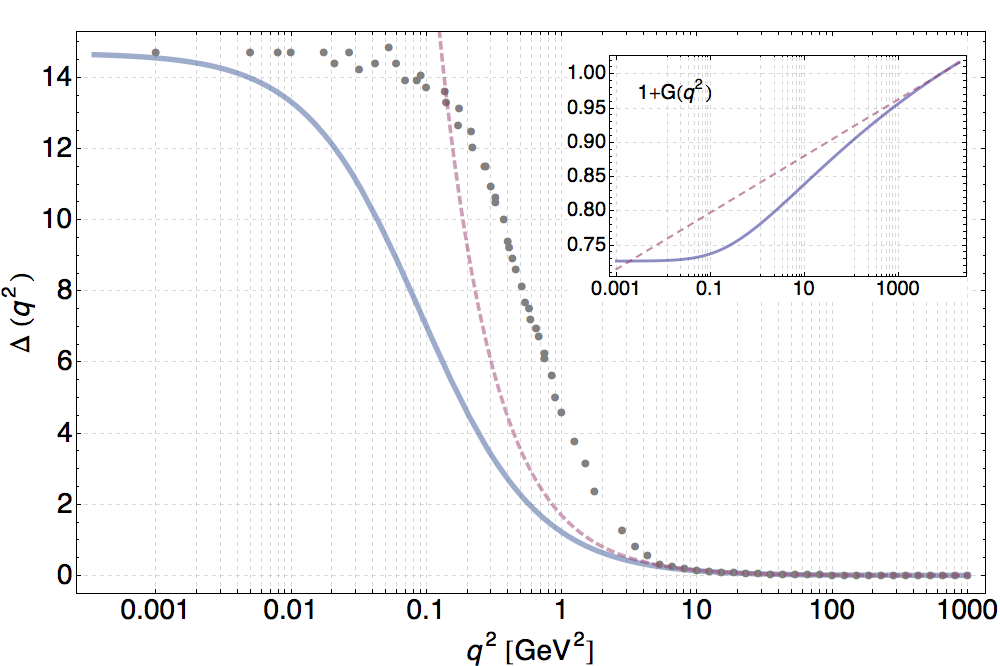} &
\includegraphics[width=8.1cm]{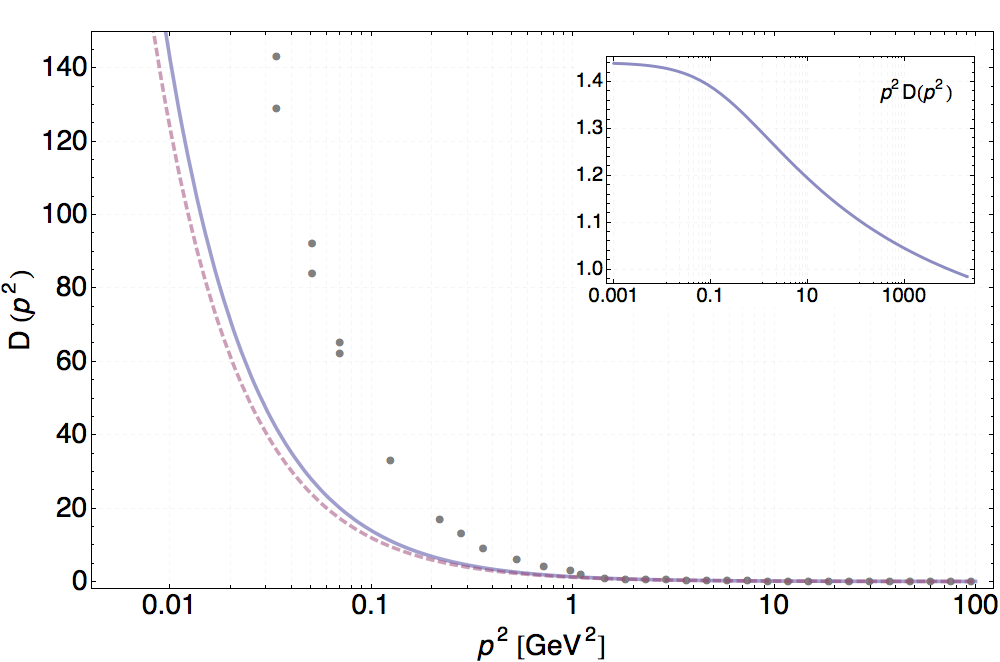}
\end{tabular}
\caption{Left panel: The gluon propagator obtained from the solution of the SDE system (blue continuous line) 
compared to the lattice data of~\cite{Bogolubsky:2007ud}; the red dashed line represents the perturbative behavior. In the inset we show the function $1+G(q^2)$ (blue continuous line) and its perturbative behavior (red dashed line).
Right panel: The ghost propagator obtained from the SDE system (blue continuous line), 
the one-loop perturbative result (red dashed line), and the 
corresponding lattice data of~\cite{Bogolubsky:2007ud}. In the inset we show the function $p^2 D(p^2)$ from the SDE.}
\label{results3}
\end{figure*}

Perturbatively  the rhs  of~Eq.(\ref{tad}) vanishes  by virtue  of the
dimensional regularization  result 
$\int_k  \frac{\ln^{n}\! k^2}{k^2}=0$ $n={0,1,2,}\dots$  
which ensures  the masslessness  of the
gluon to all orders. However, non-perturbatively $\Delta^{-1}(0)$ does
{\it not} have to vanish,  provided that the 
quadratically divergent integrals defining it 
can be properly regulated and made finite, {\it without} 
introducing counterterms of the 
form $ m^2_0 (\Lambda^2_{\chic{\mathrm{UV}}}) A^2_{\mu}$, 
which are forbidden by the local gauge invariance 
of the  fundamental  QCD Lagrangian.
It turns out that this is indeed possible:
the divergent integrals can be regulated by 
subtracting appropriate combinations of ``dimensional regularization zeros''.
Specifically, as we have verified explicitly and as can be clearly seen in 
Fig.\ref{results3} (left panel), for large enough $k^2$  
the $\Delta(k^2)$ goes over to its 
perturbative expression, to be denoted by 
$\Delta_{\rm pert}(k^2)$; it has the form  
$\Delta_{\rm pert}(k^2)=\sum_{n=0}^N a_n\frac{\ln^{n} k^2}{k^2}, $
where the coefficient $a_n$ are known from the perturbative 
expansion. For the case at hand, 
measuring $k^2$ in $\mbox{GeV}^2$, using 
$\mu \approx 100 \,\mbox{GeV}$ and $\alpha_s(\mu)=0.1$,
after inverting and re-expanding the $\Delta^{-1}(k^2)$ given below Eq.(\ref{gg}),
we find $a_0 \approx 1.7$, $a_1 \approx -0.1 $, $a_2 \approx 2.5\times 10^{-3}$.
Then, subtracting $\int_k \Delta_{\rm pert}(k^2)=0$
from both sides of Eq.(\ref{gltad}), we obtain the 
regularized $T^{\rm reg}_g$ given by ($k^2=y$) 
\bea 16\pi^2 T^{\rm reg}_g &=& \frac{15}{4} \int_0^s\!\!dy\ y
\left[   \Delta(y)    -\Delta_{\rm pert}(y)\right]
\nonumber\\   
&-& \frac{3}{2}     
\int_0^s\!\!dy\ y^2\left[\Delta^2(y)-\Delta^2_{\rm pert}(y)\right]\,.
\label{regtad}
\eea
A similar procedure can be followed for $T_c$ (see below).
The obvious ambiguity of the regularization described above is 
the choice of the point $s$, past which the two 
curves, $\Delta(y)$ and $\Delta_{\rm pert}(y)$, are 
assumed to coincide. 

Ideally, one should then:  
({\it i}) solve the system of integral equations
under the boundary condition 
$\Delta(0) =C $, where $C$ is an arbitrary positive 
parameter; ({\it ii}) substitute the 
solutions for $\Delta(q)$ and $D(q)$
in the (regularized) integrals on the rhs
of (\ref{tad}), together with the  obtained value for $G(0)$,
and denote the result by $\Delta^{-1}_{\rm reg}(0)$;
({\it iii}) check that the self-consistency requirement \mbox{$\Delta^{-1}_{\rm reg}(0)=C^{-1}$} is satisfied; 
if not, ({\it iv}) a new $C$ must be 
chosen and the procedure repeated.
In practice, due to the aforementioned ambiguity, 
we cannot pin down $\Delta(0)$ completely,
and we will restrict ourselves to 
providing a reasonable range for its value.

We have solved the system for a variety of initial values 
for $C$, ranging between \mbox{$1 -50\, \mbox{GeV}^{-2}$}, 
and obtained from (\ref{regtad})
the  corresponding $\Delta^{-1}_{\rm reg}(0)$.
On physical grounds one  
does not expect the perturbative expression $\Delta_{\rm pert}(k^2)$
to hold below \mbox{$5 - 10\,\mbox{GeV}^2$}, and therefore, 
when computing $\Delta^{-1}_{\rm reg}(0)$,
$s$ should be chosen around that value. 
For values 
of  $C$ between \mbox{$10-25\, \mbox{GeV}^{-2}$} 
 the corresponding $\Delta^{-1}_{\rm reg}(0)$ 
can be made equal to $C^{-1}$
by choosing values for $s$ within that (physically reasonable) range. 
For example, for \mbox{$C=14.7 \,\mbox{GeV}^{-2}$}, the 
value of the lattice data at the origin, 
we must choose $s \approx 10 \,\mbox{GeV}^2$. 
The solutions  for $\Delta(q)$, $D(p)$, and $1+G(q)$ 
obtained for that special choice, \mbox{$C=14.7 \,\mbox{GeV}^{-2}$}, 
are shown in Fig.\ref{results3}.
In order to enforce the equality 
\mbox{$\Delta^{-1}_{\rm reg}(0)=C^{-1}$} for higher values of $C$
one must assume the validity 
of perturbation theory uncomfortably deep into the 
infrared region;
for example, for $C=50 \,\mbox{GeV}^{-2}$ 
one must choose $s$ below \mbox{$1\,\mbox{GeV}^2$}.
We emphasize that    
the non-perturbative transverse gluon propagator, being finite in the IR, is 
automatically less singular than a simple pole, thus satisfying 
the corresponding Kugo-Ojima (KO) confinement criterion~\cite{Kugo:1979gm}, essential for ensuring 
an unbroken color charge in QCD~\cite{Fischer:2006ub}.
Note that for \mbox{$q^2\le 10\, \mbox{GeV}^2$}
both gluon propagators (lattice and SDE)
shown in Fig.\ref{results3} 
may be fitted very accurately using a unique functional form,  
given by \mbox{$\Delta^{-1}(q^2)= a+b\,(q^{2})^{c-1}$}. Specifically,
[measuring $q^2$ in $\mbox{GeV}^2$ and the $\chi^2$ per degrees of freedom],
the lattice data are 
fitted by \mbox{$a=0.07$}, \mbox{$b=0.15$},
and \mbox{$c=2.54$} (\mbox{$\chi^2\sim10^{-2}$}), 
while our SDE solution  is 
described setting \mbox{$a=0.07$}, \mbox{$b=0.77$}, and 
\mbox{$c=2.01$} (\mbox{$\chi^2\sim10^{-4}$}).

 
Let us now consider the ghosts. 
The $D(p^2)$ obtained from the ghost SDE diverges
at the origin, in qualitative agreement with the lattice data. 
From the SDE point of view, 
this divergent behavior is due to the fact that we are working in the LG 
and the vertex ${\gb}_{\nu}$ employed contains no $1/p^{2}$ 
poles, as suggested 
by previous lattice studies~\cite{Cucchieri:2004sq}.
The rate of divergence of our solution 
is particularly interesting, 
because it is related to the KO confinement criterion
for the ghost~\cite{Kugo:1979gm}, according to which 
the non-perturbative ghost propagator (in the LG) 
should be more singular in the infrared than a simple pole.
Motivated by this, we proceed to fit the 
function \mbox{$p^2 D(p^2)$} [see inset in right panel of Fig.\ref{results3}].
First we use a fitting function of the form \mbox{$p^2 D(p^2)=c_1 (p^{2})^{-\gamma}$} 
($p^2$ in $\mbox{GeV}^2$); a positive $\gamma$ 
would indicate that the SDE solution satisfies the KO  criterion.
Our best fit, valid for $p^2 \leq 10$, 
gives the values \mbox{$\gamma=0.02$} and \mbox{$c_1=1.30$}, which lead to a
\mbox{$\chi^2\sim 10^{-3}$}.
Interestingly enough, an even better fit may be obtained using a 
qualitatively different, physically motivated functional form, namely 
\mbox{$p^2 D(p^2)=\kappa_1 -\kappa_2\ln(p^2+\kappa_3)$} 
(with $\kappa_3$ acting as a gluon ``mass'').
Our best fit, valid for  the same range, 
gives \mbox{$\kappa_1=1.3$}, \mbox{$\kappa_2=0.05$}, and \mbox
{$\kappa_3=0.05$}, with  \mbox{$\chi^2\sim10^{-6}$}.
This second fit suggests that $p^2 D(p^2)$ reaches a finite (positive) 
value as $p^2\to 0$. 
Even though not conclusive, our fitting analysis seems to favor a 
ghost propagator displaying no power-law enhancement, 
in agreement with recent results presented in ~\cite{Boucaud:2008ky}; 
clearly, this question deserves further study.


Turning to the tadpole contributions $T_c$ of (\ref{ghtad}), the subtraction of 
 $0 = \int_k k^{-2}$
regularizes $T_c$, yielding a rather suppressed
finite value for $T^{\rm reg}_c$.
For example, using the first ghost fit, we get \mbox{$(s^{\prime}\approx 1 \,\mbox{GeV}^2)$} 
 \bea     
16\pi^2 T^{\rm reg}_c &=& -2 \int_0^{s'}\!\!\!\!\!dy 
\left[yD(y)-1\right]+\int_0^{s'}\!\!\!\!\!dy 
\left[y^2D^2(y)-1\right]\nonumber
\nonumber\\
&\sim & - 2 \gamma^2 s'\ln s'\,,   
\eea
which is numerically negligible.

{\it Discussion} --
The present work has focused  on the derivation 
of an infrared  finite gluon propagator
from a gauge-invariant
set of SDEs for pure QCD in the
LG,  and  its comparison  with  recent  lattice  data.  Following  the
classic  works of~\cite{Jackiw:1973tr},  the finiteness  of  the gluon
propagator  is   obtained  by   introducing  massless  poles   in  the
corresponding    three-gluon    vertex.    The   actual    value    of
$\Delta^{-1}_{\rm reg}(0)$  has been treated as a  free parameter, and
was  chosen to coincide  with the  lattice point  at the  origin.  The
curves  shown in  Fig.\ref{results3} were  then  obtained dynamically,
from  the  solution  of  the  SDE  system, for  the  entire  range  of
momenta. Comparing  the solution for the gluon propagator 
with the lattice data  we see
that, whereas their asymptotic  behavior coincides  (perturbative limits),
there is  a discrepancy of about  a factor of 2-2.5  in the intermediate
region of  momenta, especially around the  fundamental QCD mass-scale
[reflected also in the different values of the two sets of fitting parameters $(a,b,c)$].
In the case of the ghost propagator the relative difference increases
as one approaches the deep infrared, given that both curves diverge
at a different rate. These discrepancies may  be accounted for by extending  
the gluon SDE to include the ``two-loop dressed'' graphs, omitted 
(gauge-invariantly) from the present analysis,  
and/or by supplying the relevant transverse parts of the vertex given in (\ref{gluonv}).
We hope to be able to make progress in this direction in the near future. 


In our opinion,  the analysis presented here, in  conjunction with the
recent lattice  data, fully corroborates  Cornwall's early description
of QCD  in terms of a dynamically  generated, momentum-dependent gluon
mass~\cite{Cornwall:1982zr}.  In this picture the low-energy effective
theory of QCD is  a non-linear sigma model, known  as massive gauge-invariant
Yang-Mills,  obtained  from   the  generalization  of  St\"uckelberg's
construction  to  non-Abelian  theories~\cite{Cornwall:1974hz}.   
This  model  admits  vortex
solutions,  with a  long-range pure  gauge term  in  their potentials,
which endows  them with a topological quantum  number corresponding to
the center  of the gauge group  [$Z_N$ for $SU(N)$], and  is, in turn,
responsible for quark  confinement and gluon screening
~\cite{Cornwall:1979hz, Bernard:1982my}. 
Specifically, center vortices of  thickness $\sim m^{-1}$,  where $m$ is
the induced mass of the gluon, form a condensate because their entropy
(per  unit  size) is  larger  than  their  action.  This  condensation
furnishes an  area law to  the fundamental representation  Wilson loop,
thus confining quarks. On the  other hand, the adjoint potential shows a
roughly linear  regime followed by string breaking  when the potential
energy is about $2m$, corresponding to gluon screening~\cite{Cornwall:1997ds}.

{\it Acknowledgments:} 
Work supported by the Spanish MEC grants FPA 2005-01678 and 
FPA 2005-00711, and the Fundaci\'on General of the UV. 
We thank Professor J.M.Cornwall for several useful comments.
%
 
\end{document}